\documentstyle[preprint,aps]{revtex}

\tighten
\begin{document}
\bibliographystyle{/usr/share/texmf/tex/latex/revtex/prsty}
\draft

\title{
Total Photoabsorption Cross Sections of A=6 Nuclei\\
with Complete Final State Interaction}

\author{Sonia Bacca$^{1,2}$, Mario Andrea Marchisio$^{1,2}$, Nir Barnea$^3$, 
Winfried Leidemann$^{1,2}$ and Giuseppina Orlandini$^{1,2}$}

\address{$^1$
Dipartimento di Fisica, Universit\`a di Trento, I-38050 Povo, Italy} 
\address{$^2$Istituto Nazionale di Fisica Nucleare, Gruppo collegato di Trento}
\address{$^3$The Racah Institute of Physics, The Hebrew University, 91904, 
Jerusalem, Israel}

\date{\today}
\maketitle

\begin{abstract}
The total photoabsorption cross sections of $^6$He and $^6$Li are
calculated microscopically with full inclusion of the six-nucleon
final state interaction using semirealistic nucleon-nucleon  
potentials. 
The Lorentz Integral Transform (LIT) method and the effective interaction 
approach for the hyperspherical formalism  are employed. While $^6$Li has
a single broad giant resonance peak, there are two well separated 
peaks for $^6$He corresponding to the breakup of the neutron halo and
the $\alpha$ core, respectively. The comparison with the few available
experimental data is discussed.
\end{abstract}

\vfill\eject
Inelastic responses of A-body systems are fundamental physical quantities, 
since they contain valuable information about the dynamical structure of the 
system. Microscopic calculations of such responses are particularly important 
to reveal fine details of the dynamics and of the reaction mechanism. However, 
calculations of final state wave functions (FSWF)
for energies with various open channels are already rather complicated for
a three-body system and presently out of reach for A $>$ 3. On the other hand 
it has been shown that the extremely complicated microscopic calculation of 
the FSWF is not necessary, since the response can be calculated with the 
LIT method \cite{ELO94}, where only a bound state like problem has to 
be solved. Though the FSWF is not calculated, final state interaction (FSI)
 effects on the response are 
taken rigorously into account. There are quite a few examples for the 
application of this approach to few-body nuclei up to A=4 (see e.g. Refs. 
\cite{ELO97a,ELO97b,Bar01,ELOT00}). As to bound state calculations there has 
been a tremendous progress for systems with A$>$4 in the last decade.
This is due not only 
to an increase of the numerical power of modern computers, but also to 
various new microscopical approaches (GFMC \cite{GFMC}, SVM \cite{SVM}, NCSM 
\cite{NCSM}, EIHH \cite{EIHH}).  

In the present work we present a microscopic calculation of an 
inelastic reaction of a six-body system. We apply the LIT method and solve the 
above mentioned bound state like problem via an expansion in hyperspherical 
harmonics (HH) within the recently developed effective interaction (EIHH)
approach \cite{EIHH}. The response of a system to real photons is a 
fundamental quantity, therefore we have chosen the total photoabsorption cross 
sections of the six-body nuclei as inelastic response to study. It is 
important to note that photoabsorption experiments with such light nuclei 
have been performed some 20-30 years ago and that the activity has not been 
carried on further, also because of missing theoretical guidance. In the last 
years there is a renewed interest, however, mainly for halo nuclei (see e.g. 
\cite{Au98}).

Since this is the first microscopic calculation of a six-body inclusive 
reaction with consideration of the complete FSI, we take central 
nucleon-nucleon (NN) 
potential models, namely the so called semirealistic potentials 
MTI-III \cite{MT69} and MN \cite{MN77}, with parameters as given in 
\cite{EIHH}. The MTI-III model contains  Yukawa-type 
potentials and has a strong short range repulsion, while the MN model  
consists of  Gauss-type potentials and   has a rather soft core. The MTI-III 
potential is fitted to the NN scattering $S$-wave  phase-shifts, $^{1}$S$_{0}$ 
and $^{3}$S$_{1}$, up to about pion threshold, whereas the MN potential  is 
fitted to low-energy  two- and three-body data. Calculations of the total 
photoabsorption cross sections of three- and four-body nuclei have already 
been performed with the LIT method with such semirealistic potentials 
\cite{ELO97b,Bar01,ELO97c}, and recently, for $^3$H and $^3$He, also 
with realistic NN and 3N forces \cite{ELOT00}. The calculations for three--body 
nuclei show that 
semirealistic potentials lead to a rather realistic description of the total 
photoabsorption cross section. Though for six--body systems the missing P--wave
interaction could play some role, we believe that the 
main features of the calculated cross sections are close to the results of a 
more realistic calculation.
~\\
The total photoabsorption cross section is given by 
\begin{equation}
\sigma(\omega)=4\pi^{2}\alpha\omega R(\omega)\,,
\end{equation}
\noindent where $\alpha$ is the fine structure constant and 
\begin{equation}
{\label{1}
R(\omega )=\int d\Psi _{f}\left| \left\langle \Psi _{f}\right| \hat{O}\left| 
\Psi _{0}\right\rangle \right| ^{2}\delta (E_{f}-E_{0}-\omega)}
\end{equation}
\noindent is the response function; $\left| \Psi_{0/f} \right>$ and 
$E_{0/f}$ denote wave function 
and energy of ground and final state, respectively, while $\hat{O}$ is the 
transition operator. For the  
photoabsorption cross section the Siegert theorem leads to 
\begin{equation}
\label{dip}
\hat{O}=\hat{D}_{z}=\sum_{i=1}^{A}\frac{\tau^{3}_{i}z'_{i}}{2}\,.
\end{equation}
Here $\tau^{3}_{i}$  and $z'_{i}$ 
represent the third component of the isospin operator and of the coordinate of 
the $i$-th particle in the center of mass frame, respectively. It is 
well known that the dipole approximation is excellent for the total 
photoabsorption cross section (see e.g. \cite{AS91}).

In the LIT method one obtains $R(\omega)$  from the inversion of the transform
\begin{equation}
L(\sigma_{R},\sigma_{I} )= \int d\omega \frac{R(\omega )}
{(\omega -\sigma _{R})^{2}+\sigma ^{2}_{I}}= \left\langle 
\widetilde{\Psi }|\widetilde{\Psi }\right\rangle \,, \label{2}
\end{equation}

\noindent where the Lorentz state $\widetilde\Psi$ is found as  
unique solution of the ``Schr{\"o}dinger-like'' equation 
\begin{equation}
\label{3}
(H-E_{0}-\sigma_{R}+i\sigma_{I})|\widetilde{\Psi}\rangle=\hat{D}_{z}|
{\Psi_{0}}\rangle.
\end{equation}
\noindent Since the right hand side of Eq. (\ref{3}) is localized and since 
there is an imaginary part $\sigma_{I}$, one has an asymptotic boundary 
condition similar to a bound state. Therefore one can apply  bound-state 
techniques for its solution. 
We expand $\Psi_{0}$ and $\widetilde{\Psi}$ in terms of the 
six-body symmetrized HH \cite{BN97}. The expansion is performed 
up to maximal values $K^0_{max}$ and $K_{max}$ of the HH
grand-angular quantum number $K$ for $\Psi_0$ and $\widetilde\Psi$, 
respectively. In case of $\Psi_{0}$ the 
basis states are constructed with the quantum numbers of the ground state. 
With a central $S$-wave interaction for $^{6}$He ($^{6}$Li) one has: 
angular momentum L=0 (0), spin S=0 (1), isospin T=1 (0) with third component 
T$_{z}$=$-$1 (0).  For  $\widetilde{\Psi}$ the basis functions 
possess the quantum numbers selected by the dipole transition: L=1 (1), S=0 
(1), T=1 and 2 (1),  T$_{z}$=$-$1 (0). 
We improve the convergence of the HH expansion using the recently developed 
EIHH approach\cite{EIHH} where the bare potential  
is replaced by an effective potential constructed via the Lee-Suzuki method 
\cite{LeeS}. In convergence, however, the same results as with the bare
potential are obtained (see Ref. \cite{EIHH}).
 
In order to evaluate the LIT we calculate the quantity $\langle 
\widetilde{\Psi }|\widetilde{\Psi}\rangle $ directly using the Lanczos 
algorithm \cite{Mar02} instead of solving  the ``Schr{\"o}dinger-like'' 
Eq. (\ref{3}). In fact, starting from the first Lanczos vector 
\begin{equation}
\label{first}
|\varphi_0\rangle={\frac{\hat{D}_{z}|{\Psi_{0}}\rangle}{\sqrt{\langle \Psi _{0}| 
\hat{D}^{\dagger}_{z}\hat{D}_{z}| \Psi _{0}\rangle}}
}
\end{equation}
\noindent and applying  recursively  the following relations  
\begin{equation}
\label{6'}
{b_{n+1}|\varphi_{n+1}\rangle = H |\varphi_{n}\rangle + a_{n}|\varphi_{n}
\rangle -b_{n} |\varphi_{n-1}\rangle}\,,
\end{equation}
\begin{equation} 
a_{n}=\langle \varphi_{n}|H| \varphi_{n}\rangle,~~ 
b_{n}=||b_{n}|\varphi_{n}\rangle||,
\end{equation}
\noindent where $a_{n}$ and $b_{n}$ are the Lanczos coefficients, one finds 
that the LIT can be written as a continuous fraction
\begin{equation}
\label{7}
{L(\sigma_{R},\sigma_{I} )=\frac{1}{\sigma _{I}}Im\frac{\langle \Psi _{0}| 
\hat{D}^{\dagger}_{z}\hat{D}_{z}| \Psi _{0}\rangle }{(z-a_{0})-
\frac{b^{2}_{1}}{(z-a_{1})-\frac{b^{2}_{2}}{(z-a_{2})-b^{2}_{3}....}}}~.}
\end{equation}
\noindent  As shown in \cite{Mar02} a rapid convergence is reached. 

For $^{6}$Li the numerator of Eq. 
(\ref{7}) can be evaluated  directly as ground state expectation value of 
long range operators (mean square charge radius $\langle r^2_{ch}\rangle$
, mean square
proton--proton distance $\langle r^2_{pp}\rangle$),
\begin{equation}
\label{oddo}
{\langle \Psi _{0}| \hat{D}^{\dagger}_{z}\hat{D}_{z}| \Psi _{0} \rangle = 
\frac{1}{3} \left[ Z^2  \left< r^{2}_{ch} \right> - \frac{Z(Z-1)}{2}  
\left< r^2_{pp}\right>  \right], 
}
\end{equation}
\noindent that converges rapidly in the EIHH approach. For $^{6}$He the 
situation is different, because there are two final isospin channels (T=1,2). 
Therefore one has to sum over both channels, replacing Eq. (\ref{7}) by
\begin{eqnarray} \label{llant}
\lefteqn{L(\sigma_{R},\sigma_{I} ) = 
          \frac{1}{\sigma _{I}}\langle \Psi _{0}| 
          \hat{D}^{\dagger}_{z}\hat{D}_{z}| \Psi _{0}\rangle }
\cr & & \times
          Im \sum_T 
          \frac{\alpha^{T}_{K_{max}}}{(z-a_{0}(T))-
          \frac{b^{2}_{1}(T)}{(z-a_{1}(T))-\frac{b^{2}_{2}(T)}
               {(z-a_{2}(T))-b^{2}_{3}(T)....}}}~,
\end{eqnarray}
\noindent where $a_{n}(T)$ and $b_{n}(T)$ are the Lanczos coefficients
for the corresponding T channels, and
\begin{equation}
\alpha^{T}_{K_{max}}=
\frac{ \sum_{HH^T}^{K_{max}} \langle \Psi_{0}| \hat{D}^{\dagger}_
{z}|  HH^{T} \rangle \langle  HH^{T} | \hat{D}_
{z}| \Psi _{0}\rangle}
{\sum_{T}\sum_{HH^T}^{K_{max}} \langle 
\Psi_{0}| \hat{D}^{\dagger}_{z}|  HH^{T} \rangle \langle  
HH^{T} | \hat{D}_{z}| \Psi _{0}\rangle } \,.
\end{equation}
The sum $\sum_{HH^T}^{K_{max}}$ runs over all the HH basis functions
with isospin $T$ and $K \leq K_{max}$. 

Now we turn to the discussion of the results. First,
in Table \ref{tb:gs1} we present various ground state properties. From the 
estimated errors one sees that a rather
good convergence is obtained. Note that the $DD$ values can be calculated in 
two ways: as ground state expectation value and  by an 
integration of $R(\omega)$. Corresponding results differ very little 
showing a good internal consistency of our calculation. 
In Fig. 1 we show the photoabsorption cross section $\sigma(\omega)$ of 
$^{6}$Li with the MTI-III and of $^{6}$He with the MN potential for various
$K_{max}$. One observes a rather satisfactory convergence. For both 
nuclei one notes that the peak heights decrease slightly with increasing 
$K_{max}$. While the low-energy $\sigma(\omega)$ is rather stable for $^{6}$Li, 
strength is shifted towards lower $\omega$ for $^{6}$He. 
From the convergence behaviour we estimate errors of less than 10 $\%$ 
and we expect that mostly peak heights and much less the general shape of  
$\sigma(\omega)$ will be affected by possibly missing contributions from higher 
$K$.

In Fig. 2 we show our final results for both NN potentials. Though the MTI-III
model leads to somewhat higher cross section peaks, one obtains rather
similar pictures of $\sigma(\omega)$ with both potentials. 
Note that there is one single giant dipole resonance peak for $^{6}$Li and 
two well separated peaks for  $^{6}$He. The low-energy $^{6}$He peak of the 
T=1 channel at $\omega$=7.5 MeV (MN) and 9.5 MeV (MTI-III) is due to the
breakup of the neutron halo. The second one, at about $\omega$=35 
MeV, corresponds to the breakup of the $\alpha$ core and has contributions 
from both T channels (about 40 $\%$ of strength due to T=2 channel). The 
$^{6}$Li cross section does 
not show such a substructure. This is probably due to the fact that the 
breakup in two three-body nuclei, $^{3}$He~+~$^{3}$H, fills the gap between 
the halo and the $\alpha$ core peaks. Note that in case of $^{6}$He a 
corresponding breakup in two identical nuclei,  $^{3}$H~+~$^{3}$H, is not 
induced by the dipole operator. We have also integrated the various cross 
sections up to 100 MeV and find the following enhancements $\kappa_{TRK}$
of the classical TRK value (59.74 NZ/A MeVmb): 0.42 ($^6$Li, MN), 0.47 
($^6$Li, MTI-III), 0.45 ($^6$He, MN), and 0.50 ($^6$He, MTI-III).  

We already mentioned that we do not expect that the semirealistic central 
$S$-wave potentials lead to realistic results in all details. In 
particular, an additional $P$-wave interaction should affect somewhat the 
low-energy  cross section. Nevertheless we think it is instructive to 
compare with experiment. In Fig. 3a the results of 
a recent $^{6}$He experiment \cite{Au98,Aupriv} are shown. The  cross section 
was extracted from the $^{6}$He Coulomb excitation using a secondary 
radioactive $^{6}$He beam. Our results have a rather similar 
shape, but are shifted by about 2-3 MeV towards higher $\omega$. We should
mention that a better agreement close to threshold is obtained in a cluster 
model description of $^6$He with an inert $\alpha$ core and two neutrons 
interacting with it via a $P$-wave potential \cite{CoF97,DaT98}. Additional 
information comes from a recent $^6$Li($^7$Li,$^7$Be)$^6$He experiment 
\cite{nak2000} where an E1 resonance of $^6$He is found at $\omega=8.5$ MeV 
(width 15 MeV). This result is not too different 
from our low-energy peak. A similar value for the excitation energy is
found in a No-Core-Shell-Model calculation with realistic NN interactions 
\cite{nav2001}.  

In case of $^{6}$Li the experimental situation is more complex. The  
semi-inclusive channel $^{6}$Li$(\gamma,\sum_n)$  measured in Ref. \cite{Ber65}
corresponds to the total $\sigma(\omega)$ only at $\omega \le 15.7$ MeV.
At higher $\omega$, channels not involving neutrons open up 
($^{3}$He~+~$^{3}$H, $^{3}$H~+~p~+d). Regarding these two channels we 
show experimental data from Refs. \cite{Jun79,Sh75}. To make the 
comparison simpler we have summed these data with those of \cite{Ber65} 
(see Fig. 3b). At low $\omega$ the comparison of 
theoretical and experimental results is rather similar to the $^{6}$He case 
with a shift of the theoretical cross section to somewhat higher $\omega$. The 
comparison does not improve with increasing $\omega$. On the other hand 
the experimental situation is not settled as one can note from the different 
results of Refs. \cite{Jun79} and \cite{Sh75}.

In the following we summarize our results briefly. We present the first 
microscopic calculation of an inelastic six-body cross section  
considering the complete six-body FSI. To this end we make 
use of the LIT method \cite{ELO94} and expand ground and Lorentz states in 
hyperspherical harmonics via the EIHH approach \cite{EIHH}. The LIT is 
calculated with the help of the Lanczos technique \cite{Mar02}. 
The calculated total photoabsorption 
cross sections of $^6$He and $^6$Li show rather different structures. While 
$^6$Li exhibits a single broad giant resonance peak, one clearly distinguishes 
two well separated peaks for $^6$He. The low-energy peak is due to the breakup 
of the $^6$He neutron halo, whereas the second peak corresponds to the breakup 
of the $\alpha$ core. A comparison with experimental data shows that the 
theoretical cross sections miss some strength at very low energies, which is 
presumably explained by the missing $P$-wave interaction in the employed 
semirealistic central $S$-wave potentials. The situation in the giant 
dipole peak region is much less clear, either because of lack of experimental 
data ($^6$He) or because existing data do not lead to a unique picture 
($^6$Li). It is evident that further experimental activities are necessary in 
order to shed more light on the six-nucleon photoabsorption cross sections. On 
the other hand further progress has also to be made in theory, in particular 
by addressing the question of the role of $P$-wave interactions in the total 
photoabsorption cross section of the six-body nuclei.

\begin{table}
\caption{\label{tb:gs1}
Binding energy ($E_B$), root mean square matter radius ($r_m$)
and ground state expectation value in Eq. (\ref{oddo}) ($DD$) for
$^6$He and $^6$Li with MN and MTI-III potentials (Coulomb force included). 
The numbers in brackets 
are an estimate of the error due to the convergence behaviour.}
\begin{center}
\begin{tabular}
{llcccc}
 Nucleus & $V_{NN}$ & $K^0_{max}$ & $E_B$ [MeV]  & $r_m$[fm] &  $DD$[fm]  
\\ \hline
$^6$He & MN & 10  &  -30.48(10) &   2.37(4)  &  2.28(15)  \\ 
       & MTI-III  & 10  &  -31.87(10) &   2.23(4)  &  1.99(10)  \\
$^6$Li & MN & 12  &  -34.90(10)  &   2.18(2)  &  1.50(2)  \\
       & MTI-III  & 12  &  -35.88(20) &   2.13(2)  &  1.51(1) 
\end{tabular}
\end{center}
\end{table}

\begin{figure}
\caption{cross section $\sigma(\omega)$ with various values of 
$K_{max}$: $^{6}$Li with MTI-III (a) and $^{6}$He with MN potentials (b).}
\end{figure}

\begin{figure}
\caption{cross section $\sigma(\omega)$ with both NN potentials: $^{6}$Li (a), 
$^{6}$He with channel T=2 and sum of T=1 and 2 (total) (b).}
\end{figure}

\begin{figure}
\caption {Theoretical and  experimental results for cross section
$\sigma(\omega)$: 
$^{6}$He (theoretical curves are convoluted with the gaussian instrumental 
response function [10,18]) (a), $^{6}$Li with experimental data from [23-25] 
(see text) (b).}
\end{figure}

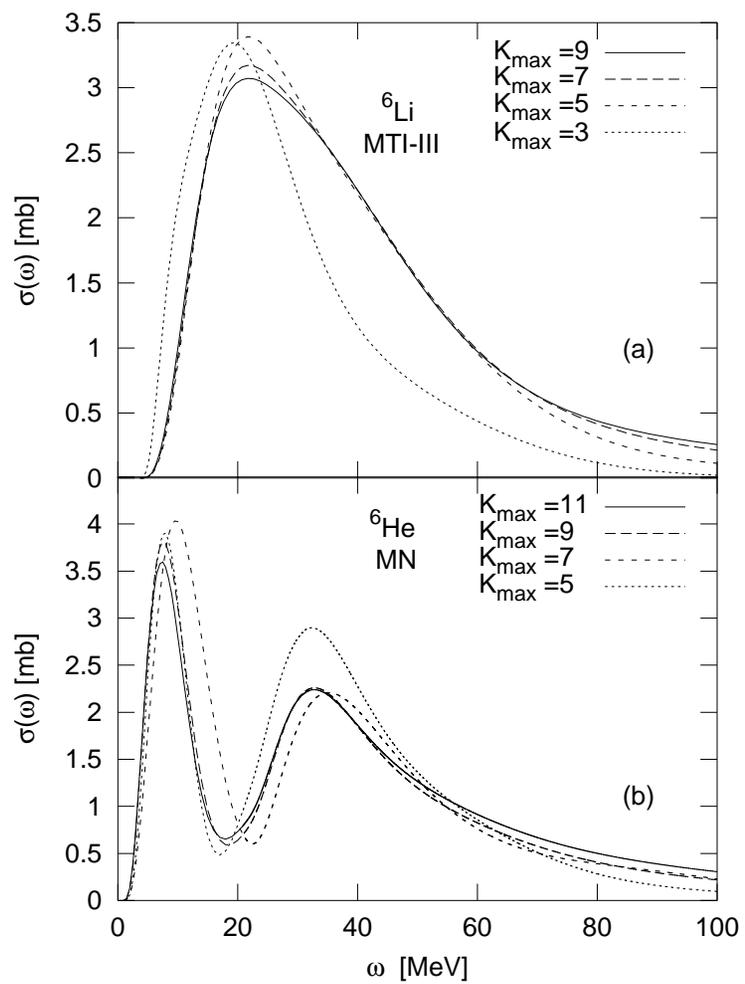
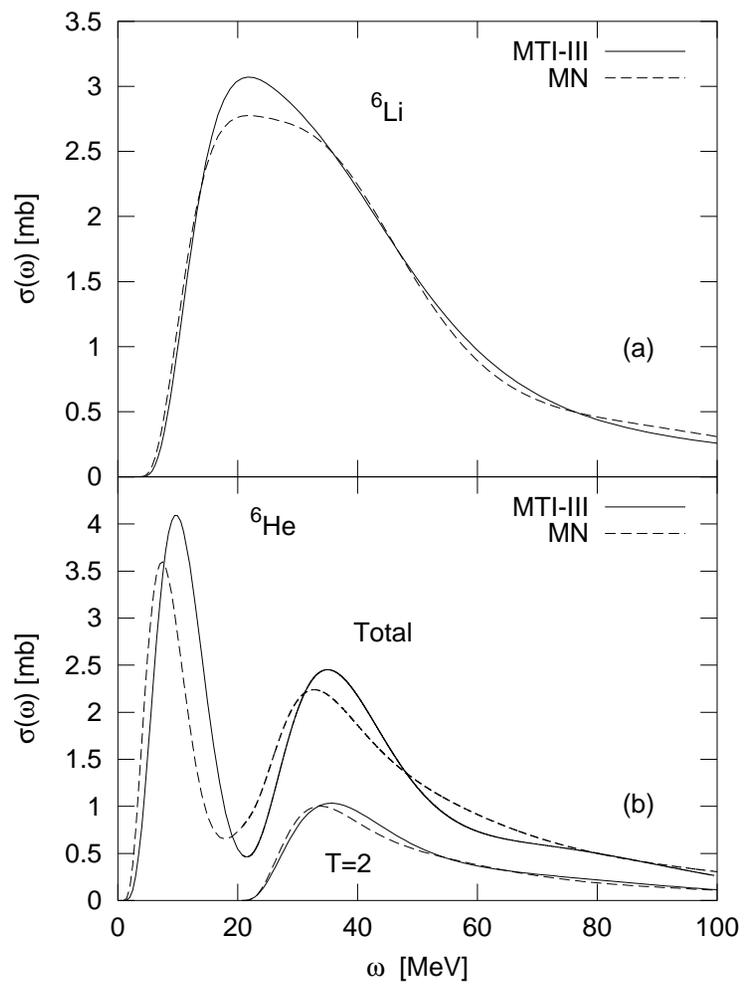
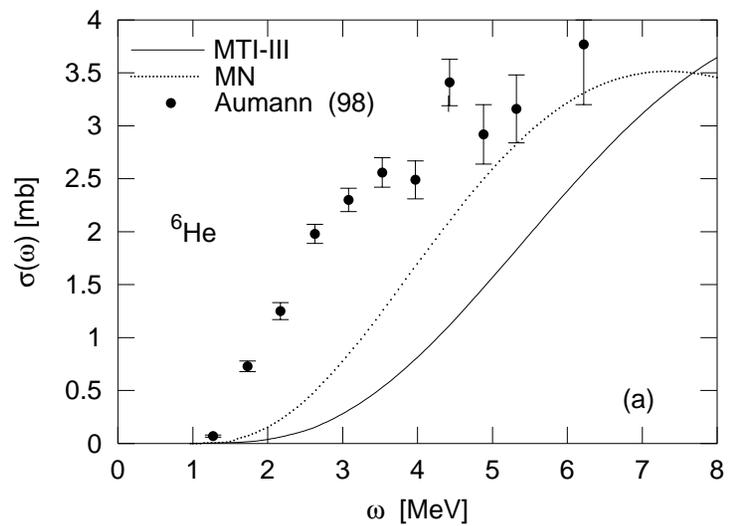
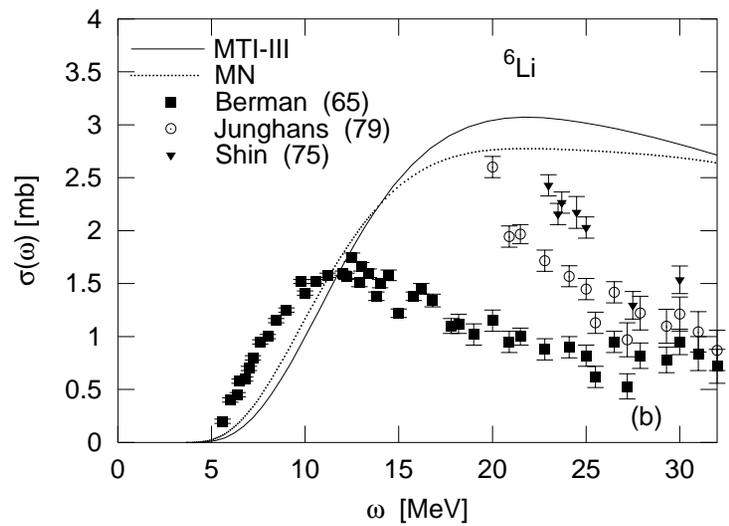
\end{document}